\documentclass[iop,twocolumn,apj]{emulateapj}
\usepackage{graphicx}
\usepackage{amsmath}

\newcommand{\kmprs}{\mbox{\rm\,km\,s$^{-1}$}}
\newcommand{\feh} {\mbox{\rm [Fe/H]}}
\newcommand{\teff}  {\mbox{$T_{\rm eff}$}}
\newcommand{\logg}  {\mbox{{\rm log}$g$}}
\newcommand{\gmr} {\mbox{$(g-r)_{\rm 0}$}}
\newcommand{\Vrot} {\mbox{$V_{\rm rot}$}}

\newcommand{\Vlsr} {\mbox{$V_{\rm LSR}$}}

\newcommand{\Mg} {\mbox{$M_{\rm g}$}}
\newcommand{\Mi} {\mbox{$M_{\rm i}$}}

\shorttitle{RHB Stars from SDSS survey}
\shortauthors{Chen et al.}

\begin{document}


\title{Metallicity and Kinematic distributions of Red Horizontal-Branch Stars from the SDSS Survey}


\author{Y.Q. Chen$^{1}$, G. Zhao$^{1}$, J.K. Zhao$^{1}$, X.X. Xue$^{1}$, W.J. Schuster$^{2}$}

\altaffiltext{1}{Key Laboratory of Optical Astronomy, National Astronomical Observatories, Chinese
Academy of Sciences, Beijing, 100012, China; cyq@bao.ac.cn.}
\altaffiltext{2}{Observatorio Astronomico Nacional, Universidad Nacional Autonoma de Mexico, Apartado Postal 877, C.P.22800 Ensenada, B.C., Mexico;schuster@astrosen.unam.mx}

\begin{abstract}
On the basis of a recently derived color-metallicity relation
and stellar parameters from the Sloan Digtal Sky Survey Data Release 7 spectroscopic survey,
a large sample of red horizontal branch candidates have been selected
to serve as standard candles.  The metallicity and kinematic
distributions of these stars indicate that they mainly originate
from the thick-disk and the halo populations. The typical thick
disk is characterized by the first group peaking at $\feh \sim -0.6$,
$\Vrot \sim 170 \kmprs$ with a vertical scale
height around $|Z| \sim 1.2$ kpc, while stars with $\feh < -0.9$ are
dominated by the halo population.  Two sub-populations of the halo
are suggested by the RHB stars peaking at $\feh \sim -1.3$:
one component with $\Vrot > 0 \kmprs$ (Halo I) shows a sign of metallicity
gradient in the $\feh$ versus $|Z|$ diagram, while the other with
$\Vrot < 0 \kmprs$ (Halo II) does not. The Halo I mainly clumps
at the inner halo with $R < 10$ kpc and the Halo II comes both
from the inner halo with $R < 10$ kpc and the outer halo with  $R > 10$ kpc
based on the star distribution in the $R$ versus $|Z|$ diagram.
\end{abstract}


\section{Introduction}

Red horizontal-branch (RHB) stars are found in many Galactic
globular clusters (GCs) with intermediate metallicity
$-1.7 < \feh < -0.3$ (see, e.g.,\ Piotto et al.\ 2002). This metallicity
range includes stars from the halo, the thick disk, the thin disk, and
probably the bulge. Due to their nearly constant absolute magnitude,
this type of stars constitutes an ideal sample to trace various
Galactic populations in the Galaxy.

Generally, the identification of RHB stars in the field
 is rather difficult just by their colors but some attempts 
 to identify this type of star had been made (Rose 1985;  Tautvaisiene
 1996; Preston 2006). However, these works did not produce a large
 sample of RHB stars (for example, a sample with star number larger
 than 100) for statistic work.
 With the aid of
Hipparcos parallaxes, Kaempf et al.\ (2005) selected a large sample
of RHB stars in the field and calculated their kinematical parameters
based on available radial velocities and proper motions. They found
two populations:  a disk one having $|Z| \sim 0.6$ kpc and a halo one
showing $|Z| \sim 4$ kpc.  This is the only work to trace the structure
of the Galaxy via a large sample of RHB stars. Due to the distance
limitation of the Hipparcos survey, these results were based on stars
in the solar neighborhood, where the majority of the disk stars are
from the thin disk with a minority from the thick disk and only a
sprinkling of halo stars. It is of high interest to trace the disk
and halo populations via RHB stars locating far away from the solar
neighborhood and well above the Galactic plane.  For this purpose,
planned or ongoing kinematically unbiased surveys, such as the
Global Astrometric Interferometer for Astrophysics (GAIA; Turon et al.\ 2005), the SDSS/SEGUE (York et al.\ 2000; Yanny et
al.\ 2009), and the LAMOST (Zhao et al.\ 2006) projects, will provide
a chance to obtain and to study a large sample of RHB stars to the
distant halo.

In this work, we attempt to investigate the properties of RHB stars and
to carry out kinematic traces of various Galactic populations using
RHB stars selected from the SDSS
spectroscopic survey. As compared with the previous study by Kaempf et
al.\ (2005), the present work has several advantages. First of all, the
SDSS survey covers quite distant regions of the Galaxy, and our study
can be extended far away from the solar neighborhood.
Second, spectroscopic data of the SDSS survey provide stellar parameters,
radial velocities, and metallicities.  This data set not only makes it
possible to identify RHB candidates more easily, but also provides the
most powerful tool to investigate the Galactic populations by a
combination of chemical and kinematical ingredients, in view of the fact
that our Galaxy evolves both chemically and dynamically. Finally, it is
expected that RHB stars selected from the SDSS survey could extend to
a much lower metallicity range than that of Kaempf et al.\ (2005),
which mainly concentrated on solar metallicity, and thus they will
contribute much to the understanding of the thick-disk and the halo
populations.

\section{Selection of RHB Stars}

RHB stars are easily seen based on the color-magnitude diagrams (CMDs)
of GCs, and statistically it is possible to pick out this type of star in the
field of the solar neighborhood with the aid of Hipparcos parallaxes.  Without
previously-known distance information for stars far from the solar neighborhood,
the selection of RHB stars in the field is rather difficult.  But the problem of
identification is much easier if the metallicity of the star is known, as shown
in Straizys et al.\ (1981).  Based on high resolution and high signal-to-noise ratio
spectra, around 30 RHB stars have been identified in the literature (e.g., Behr et
al.\ 2003; Carney et al.\ 2003, 2008).  In order to enlarge this sample, RHB stars
have been identified from the SDSS low resolution (R $\sim$ 1800) spectroscopic
survey.  The SDSS
Data Release 7 (DR7; Abazajian et al.\ 2009) provides photometry and
spectroscopic metallicities for a large sample of high-latitude stars, allowing
the identification of a large sample of RHB stars in the field.

\subsection{The Metallicity-Color Relation}

As noticed in Chen et al.\ (2009), the $\gmr$ values of RHB stars in
metal-poor GCs with $\feh=-1.6$ dex, such as Pal 3, NGC\,7006, and M3,
are about 0.25-0.30 mag and extend to the red, reaching 0.60 in M71 with
$\feh=-0.8$ and to 0.98 in NGC\,6791 with $\feh=+0.4$. In fact, there
is a correlation between $\gmr$ and $\feh$ for RHB stars in clusters.
In order to obtain a quantitative relation, we perform a linear
regression to the data in Chen et al.\ (2009), which gives\\
$\gmr = 0.343 (\pm0.039) \feh+0.829$. \\
%% (\sigma=0.069)$ \\
The scatter around this relation is approximately 0.07 mag.  Figure~1
shows the peaking $\gmr$ (and its error) versus $\feh$ for each
cluster and a linear relation is obtained by fitting to these points.

\begin{figure}[bt]
\includegraphics[scale=1.00]{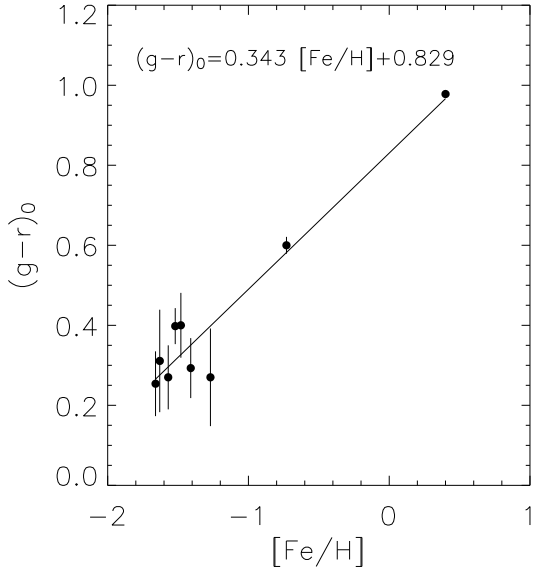}
\caption{$\gmr$--$\feh$ diagram for clusters in Chen et al.\ (2009).
The solid line indicates a linear fitting to the data.}
\end{figure}

Assuming that RHB stars in the field follow the same $\gmr$ and $\feh$
relation as these clusters, the first step is to apply this
color-metallicity relation to all stars in SDSS-DR7 where spectroscopic
metallicities and photometric colors are available in the published
catalog. In the present work, the interstellar reddening values $E(B-V)$
are based on the extinction map of Schlegel et al.\ (1998) and we adopt
the reddening corrections for $ugriz$ colors from Fukugita et
al.\ (1996).  In the selection, a deviation in $\gmr$ to the above relation
$\delta \le 0.15\ mag$  is used to select stars; this value is adopted by
taking into account an error of 0.2 dex in $\feh$ in SDSS-DR7, an
error of 0.1-0.2 $mag$ in the reddening uncertainty, and an approximate
$\gmr$ color extension of 0.1 $mag$ for RHB stars in cluster.
Then, we select stars with signal-to-noise ratios of their spectra
larger than 10 and metallicity larger than $\feh > -2.0$
so that their radial velocities and stellar parameters
can be reliably determined from the SDSS-DR7 catalog. Finally,
an upper cutoff at $g=20\,mag$ is necessary because the proper
motions from the USNO survey become seriously contaminated by
misidentifications toward faint magnitudes according to Fuchs et al.\ (2009).
But the effective cutoff in magnitude of our final sample 
is around $g=18 - 18.5\,mag$ and the saturation flag
in the SDSS-DR7 catalog is used to avoid stars whose magnitudes
suffer from saturation effects.
With these criteria, we obtain our preliminary sample of $\sim 120,000$ stars.

\subsection{The $\teff$ versus $\logg$ diagram}

Using the above $\gmr$--$\feh$ relation, blue horizontal branch (BHB) and RR Lyrae stars are
excluded at the bluer colors, and red giant branch stars at the red
limit, for a given metallicity.  However, the sample is significantly
contaminated by main-sequence stars (MS), turnoff (TO), and
subgiant (SG) stars with the same metallicity; these can be
distinguished if their luminosity or gravities are known. In the CMD
of the clusters in Chen et al.\ (2009), at a given metallicity, RHB
stars have brighter absolute magnitudes
than those of MS/TO/SGB stars despite their similar
colors, and the separation
between RHB and MS/TO/SGB stars in absolute magnitude at a given color
becomes larger as the metallicity decreases.  In view of this, stellar
parameters, i.e., temperature, gravity
and metallicity, taken from the SDSS-DR7 catalog,
can be used to identify RHB stars.
The $\teff$--$\logg$ diagram for the first 20000
stars in the preliminary sample and the
corresponding contour map is shown in Figure~2.
This figure shows that MS/TO/SG
and RHB stars can be distinguished with RHB stars clumping over the range
of 4500--5900\,K in temperature and 1.8--3.5 dex in $\logg$, which are
the selection criteria for our sample stars. In order
to check these temperature and gravity ranges for RHB stars, we have
compiled stellar parameters of previously identified RHB stars in the
field and in clusters based on high resolution spectral analyses in the
literature.  Most works (Behr et al.\ 2003; Carney et al.\ 2003, 2008)
agree with the ranges of 4600--6200\,K in temperature and of 2.0--3.3
dex in $logg$ for RHB stars, which are consistent with the defined
ranges of RHB stars in Figure~2.

\begin{figure}[bt]
\includegraphics[scale=1.00]{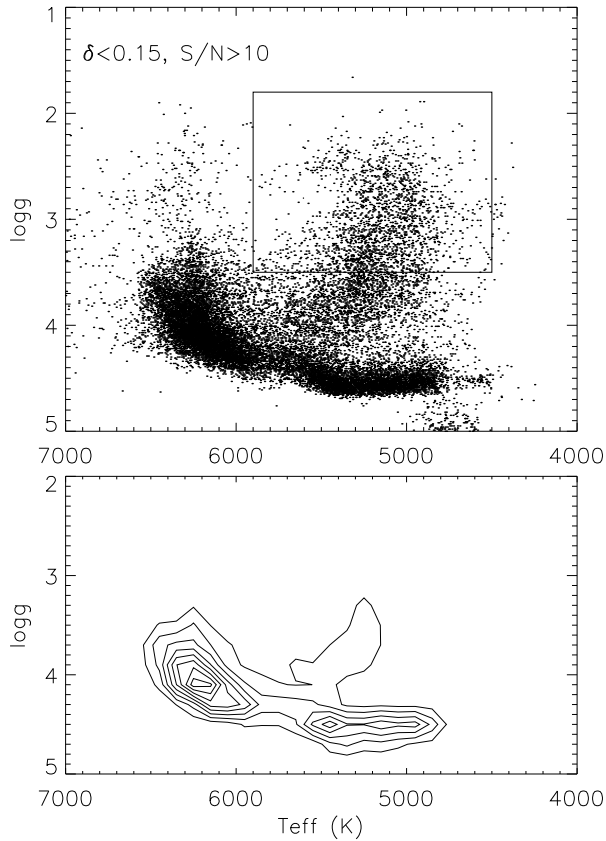}
\caption{$\teff$ - $\logg$ diagram and its contour map for
 the first 20000 stars in the preliminary sample. The box indicates our
selected RHB candidates.}
\end{figure}

\section{Stellar distances and space velocities}

\subsection{Distances, proper motions, and radial velocity}

Stellar distances have been calculated using the absolute magnitude-metallicity
relation of $\Mg=0.492\feh+1.39$ derived by Chen et al.\ (2009). We have
compared them with distances derived from the $i$ magnitude and $\Mi-\feh$
calibration derived in Chen et al.\ (2009), and they give similar distances.
The radial velocities and
proper motions are based on values provided by the SDSS-DR7 catalog. Note that
the proper motions in the SDSS-DR7 catalog have been corrected for the
systematic error noticed by Munn et al.\ (2008).  Finally, only stars
with high quality data in photometry ($\sigma_g < 0.1$, $\sigma_i < 0.1$ mag),
in proper motions ($\sigma_{\mu} < 3 mas\ yr^{-1}$), and in radial velocity
($\sigma_ {RV} < 10 km\ s^{-1}$) are included in the sample.

\subsection{Spatial velocity}

Stellar positions (X,Y,Z) and heliocentric space velocity (U,V,W) are
calculated following the same procedures as in Chen et al.\ (2000).
We use a left-handed system in which U is positive toward the Galactic
anti-center. The solar motion of (U,V,W)$_\odot$ = (7.5, 13.5, 6.8) $km\ s^{-1}$ 
with respect to the LSR from Francis \& Anderson (2009) and the local standard
of rest (LSR) velocity of
220 km/s (corresponding to $V_{hel} \sim -225 km\ s^{-1}$) are adopted
in the present work.

The errors in spatial velocity come from the uncertainties of radial
velocity, proper motion and distance.  The error in radial velocity of
3 km/s corresponds to a spatial velocity error of 5 $km\ s^{-1}$. The distance
error of 10\%, estimated from an error in absolute magnitude of 0.2 $mag$,
corresponds to a velocity
error of 5-15 $km\ s^{-1}$.  The largest error comes from uncertain proper motions.
As shown in Ivezic et al.\ (2008), a random error of 3 $mas\ yr^{-1}$ in proper
motions will lead to a velocity error of about 15 $\kmprs$ at a distance of
1 kpc and about 80 $km\ s^{-1}$ at a distance of 5 kpc.  Although the total error
in spatial velocity is quite large, the data still represent usable
measurements for a large sample because the systematic errors are much
smaller ($\sim$10 $km\ s^{-1}$ at a distance of 7 kpc) according to Ivezic et al.\ (2008).

In principle, a total velocity larger than $600 \kmprs$ is unreasonable,
or indicates stars unbound to our Galaxy. Such stars are usually
excluded from the final sample (e.g., Klement et al.\ 2009). In the
present work, we keep them in our sample for further check, but we have
checked that the main results are generally the same when we excluded
them from the sample.
With this in mind, we avoid to overexplain the results based
on the spatial velocities only and instead
statistical results based on metallicity and
distance to the Galactic plane ($|Z|$) are more favorable in
the following analysis.

The final sample
includes 5391 stars for further analyses. Figure~3 shows the distributions of distances, $g$ magnitude,
reddening, and galactic latitudes of the final sample.
Their identifications, stellar
parameters, and spatial velocities are published electronically and a sample
table consisting of the first 10 stars is presented in Table 1.

\begin{figure*}[bt]
\includegraphics[scale=1.00]{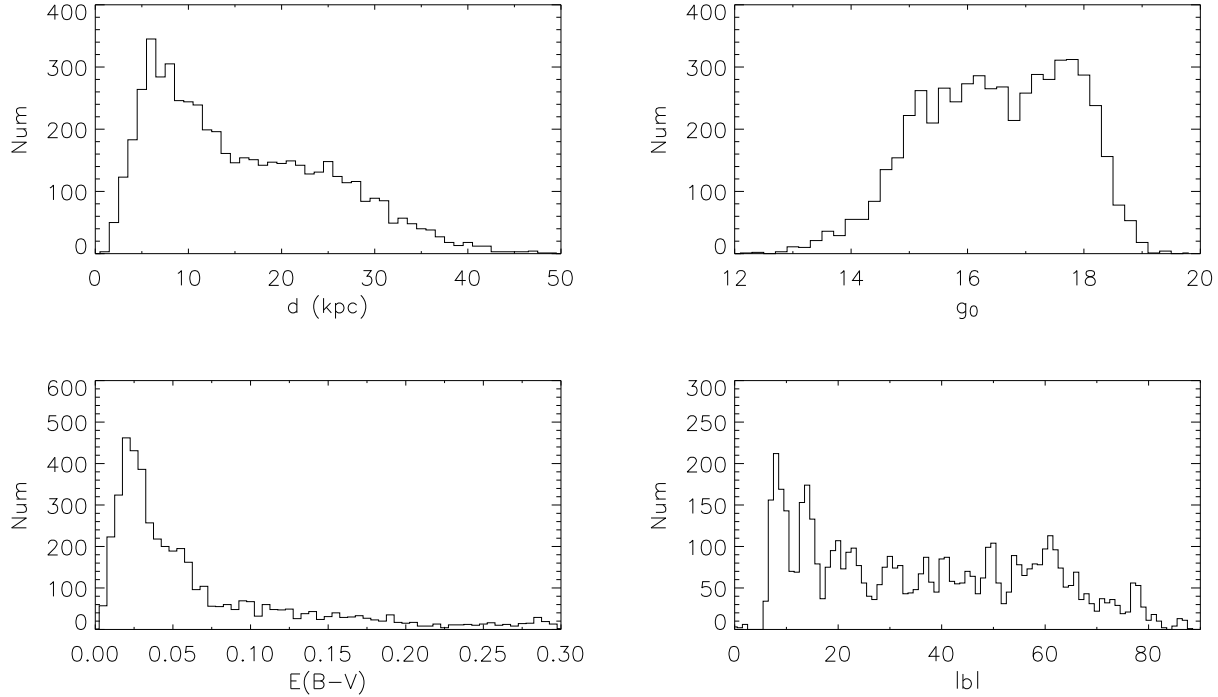}
\caption{Dstributions of distances, $g_0$ magnitude,
reddening and galactic latitudes of the final sample.}
\label{sample}
\end{figure*}

\begin{table*}
\caption{Identifications, stellar
parameters, coordinates, and spatial velocities of the first ten stars in the
final sample are presented.}
\label{tb:Mx} \setlength{\tabcolsep}{0.08cm}
\begin{tabular}{rrrrrrrrrrrrrr}
\noalign{\smallskip}
\hline
plate & mjd    & fiber & $\feh$ & Z & d & $g_0$ &  RA & DEC & l &  b & U  & V & W \\
\noalign{\smallskip} \hline
      &        &       &        & (kpc) & (kpc) & (mag) &  (deg)& (deg) & (deg) &  (deg) & ($\kmprs$)& ($\kmprs$)& ($\kmprs$) \\
\noalign{\smallskip} \hline
  408&51821 & 528 &  -0.40 & -1.76  &2.26 &12.9  & 40.40  & 1.24 &170.43 &-51.18 & -12.2 & -36.0 &  14.7 \\
  573&52325 & 351 &  -0.48 &  3.44  &4.99 &14.6  &150.36  & 4.35 &234.60 & 43.54 & 192.4 & -30.2 & -53.8\\
  580&52368 &  25 &  -0.47 &  4.70  &5.70 &14.9  &166.10  & 4.22 &250.11 & 55.70 & 147.6 & -43.0 &  24.1\\
  934&52672 & 473 &  -0.26 &  2.77  &4.51 &14.5  &131.64  &35.58 &187.48 & 37.86 & -27.7 &-137.4 & -99.4\\
 1246&54478 & 542 &  -0.54 & -0.94  &3.90 &14.1  & 86.88  & 0.52 &205.03 &-13.94 & 102.5 &   9.8 & -22.4\\
 1247&52677 & 590 &  -0.26 & -0.61  &2.78 &13.5  & 88.28  & 0.60 &205.62 &-12.66 &  16.8 & -42.5 &  23.8\\
 1662&52970 & 136 &  -0.33 & -0.61  &4.05 &14.3  &351.21  &51.95 &109.63 & -8.67 &-132.2 & -45.1 &  55.0\\
 1910&53321 &  27 &  -1.25 & -5.46  &6.37 &14.8  &345.85  &-9.63 & 62.27 &-59.09 & -63.8 &-622.0 &-351.3\\
 1910&53321 & 276 &  -0.73 & -4.64  &5.50 &14.7  &343.82  &-9.62 & 59.97 &-57.50 & 304.3 &-360.5 &-199.8\\
 1960&53289 & 416 &  -0.46 & -1.51  &3.31 &13.7  &322.40  &12.24 & 65.01 &-27.19 & -84.4 & -99.1 &  38.3\\
\hline
\noalign{\smallskip}
\end{tabular}
\end{table*}

\section{Results and Discussions}
\subsection{The metallicity and kinematical distributions}

By choosing a single stellar type, RHB stars, it is interesting to
investigate the metallicity distribution and compare our results with
those from other types of stars, e.g.,\ MS or TO 
stars.  In Figure~4, a histogram of our sample shows
two peaks in the metallicity distribution with a division at
$\feh \sim -0.9$ dex.  Two Gaussian fits to the distribution show a
metal-poor sub-population at $\feh \sim -1.3$ dex, and a metal-mild
sub-population at $\feh \sim -0.6$ dex.  As compared with the result
from MS stars within 4 kpc of the Sun in the SDSS I and II
survey by Carollo et al.\ (2010), we find that the two peaks in the
metallicity distribution are quite similar. They suggested that stars
in the metal-mild peak belong to the disk population (mainly the
thick disk), while stars in the metal-poor peak belong to the halo
population. However,
there are some differences in star numbers for each population between
RHB stars in the present work and MS stars in Carollo et al.\ (2010).
In our work, the metal-poor component, peaking at $\feh \sim -1.3$,
outnumbers the metal-mild component, peaking at $\feh \sim -0.6$, which
is opposite to the result in  Carollo et al.\ (2010), where the metal-mild
component surpasses the metal-poor component.  This is easily understood
considering that their sample of stars is limited to $d < 4$ kpc from the
Sun, and thus the disk population contributes significantly to their
sample, while RHB stars in our sample have brighter absolute magnitudes
and extend to more distant regions of the Galaxy where the halo dominates.

The variations in the metallicity distribution with
different distances to the Galactic plane ($|Z|$) are shown in Figure~5,
where the metallicity peak shifts from $\feh \sim -0.4$ dex for
 $|Z|<1.5$, $\feh \sim -0.6$ dex for $1.5 \leq|Z|\le5$
to $\feh \sim -1.3$ dex for $|Z|>8$. At $5 \leq|Z|\le8$, the
star numbers between the two
components are comparable. Note that the result
in Figure~20 of Carollo et al.\ (2010) shows the shift of the
metallicity peak from $-0.6$ at 0 kpc $ < |Z| < 1$ kpc,
to $-1.3$ at 3 kpc $ < |Z| < 4$ kpc, reaching to $-1.6$ at 6 kpc$ < |Z| < 7$ kpc and
to $-2.2$ at $|Z| > 9$ kpc.
Therefore, the exact $\feh$ peaks at a given $|Z|$ between
Carollo et al.\ (2010) and the present work are somewhat
different due to the selection of different spectral types of
stars in the sense that MS/TO stars in their sample include
significant contributions from the thin disk and the halo with $\feh < -1.6$,
while our sample of RHB stars has the largest contribution from the thick disk,
the minority contribution from the halo and nearly negligible contribution from
the thin disk.

\begin{figure}[bt]
\includegraphics[scale=1.00]{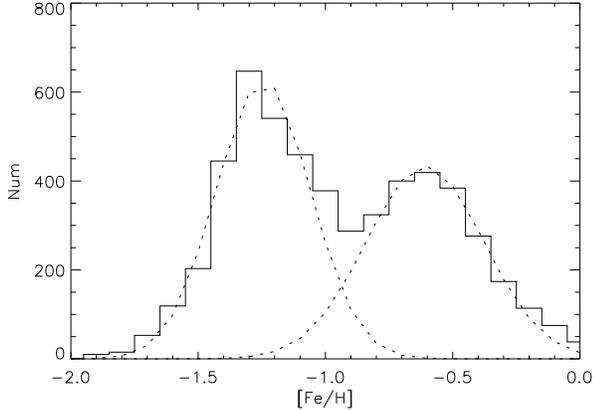}
\caption{Metallicity distributions of RHB stars. The dotted lines
show the Gaussian fits to the two populations separated by
the metallicity division at $\feh \sim -0.9$.}
\end{figure}

\begin{figure}[bt]
\includegraphics[scale=0.90]{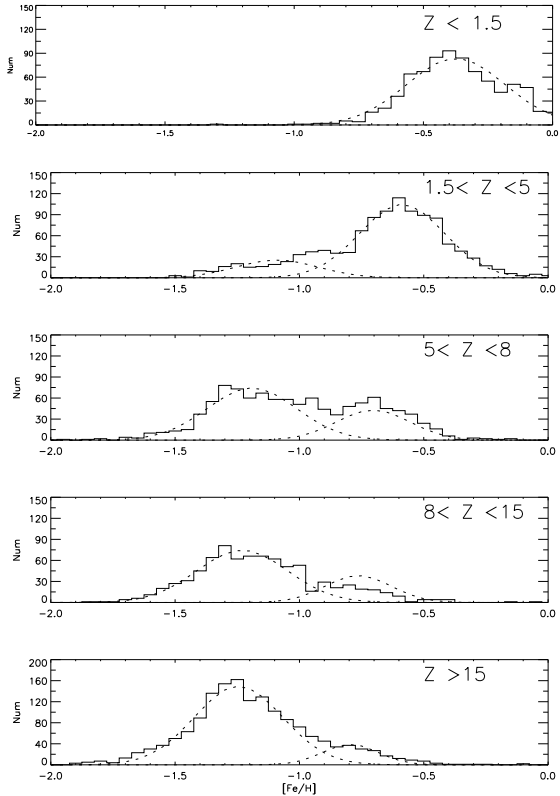}
\caption{Metallicity distributions of RHB stars at different $|Z|$.}
\end{figure}

\begin{figure}[bt]
\includegraphics[scale=0.90]{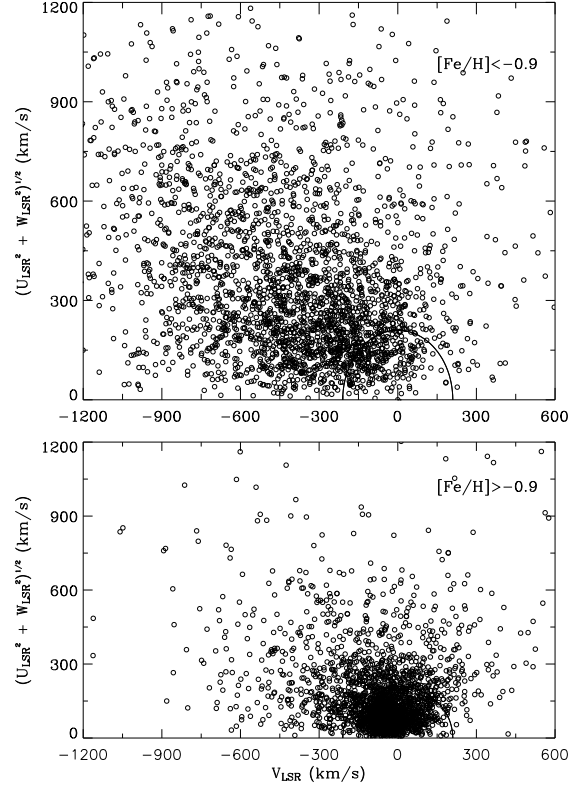}
\caption{Toomre diagrams of RHB stars for $\feh > -0.9$
and  $\feh < -0.9$. The circles indicate the division between
the thick disk and the halo following Nissen \& Schuster (2010).}
\end{figure}

\begin{figure}[bt]
\includegraphics[scale=0.90]{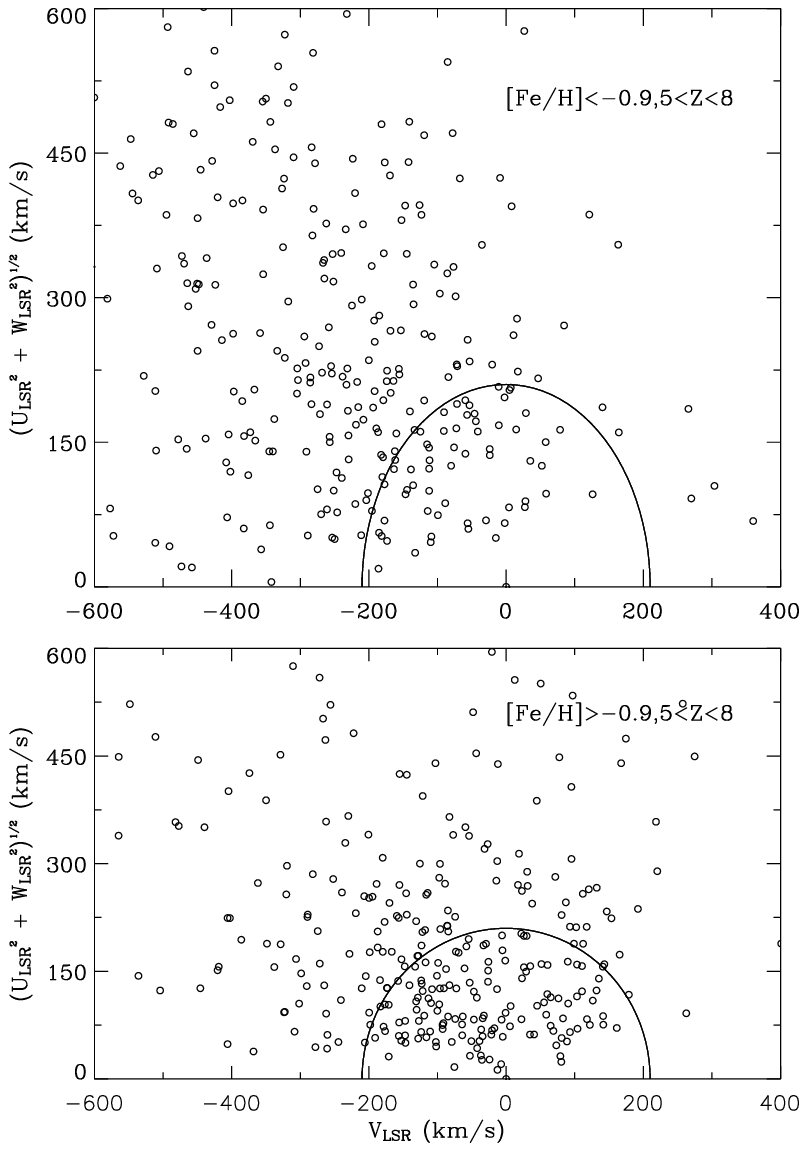}
\caption{Toomre diagrams of RHB stars with $5 < |Z|<8$ kpc for $\feh > -0.9$
and  $\feh < -0.9$, respectively. The symbols are the same as Figure~6.}
\end{figure}

In order to investigate the origins of the two populations in the
metallicity distribution for RHB stars in the present work, the Toomre
diagrams for stars with $\feh > -0.9$ and  $\feh < -0.9$ are shown in
Figure~6 where the solid circle indicates the division between the
thick disk and the halo with $V_{tot} = 210 \kmprs$
following Nissen \& Schuster (2010). Note that there is a
significant number of stars with
a total velocity larger than $600 \kmprs$ in these figures due to
the very large errors in the determination of space velocities
when the distance of the star is larger than 10 kpc. Therefore,
we do not attempt to draw any firm conclusion from the kinematical
data, but the statistical argument for dividing stars into different
groups may be right.

It is clear that the metal-mild component of RHB stars clumps within the
thick-disk region with some extension to the thin-disk region, while the
majority of stars in the metal-poor component are located in the halo
region.  This result is consistent with that of Carollo et al.\ (2010)
in the sense that the metal-mild component, peaking at $\feh \sim -0.6$,
mainly originates from the thick disk and the metal-poor component,
peaking at $\feh \sim -1.3$, generally belongs to the halo.
In Figure~7, it is interesting to compare the kinematic diagrams between the
two metallicity peaks for stars with 5 kpc $ <|Z|< 8$ kpc in Figure~6 (the third panel).
Again, it shows that the metal-poor component at $\feh \sim -1.3$
mainly has the halo kinematics and the metal-mild component at $\feh \sim -0.6$
has the thick disk kinematics. In the following sections, we
separate the stars into two parts with  $\feh > -0.9$ and  $\feh < -0.9$,
respectively.

\subsection{The $\feh$ versus $|Z|$ diagram}
Figurere~8 shows the distribution of 5391 RHB stars in the
$\feh$ versus $|Z|$ diagram with the pink color corresponding to a high
density, green to middle density, and black to the lowest density.
The division between
the thick disk and the halo seems to be quite clear in
the $\feh$ versus $|Z|$ diagram as clearly shown in Figure 8, where the edge
of the thick disk could be as high as $|Z| \sim 8$ kpc at $\feh \sim
-0.6$ and reduces to $|Z| \sim 2$ kpc at $\feh \sim -1.5$.
Figurere~9 (upper panel) shows individual stars in this diagram.
It seems that there is a metallicity gradient in the vertical direction of the
Galaxy, but the scatter around the gradient is quite large.  The $|Z|$
range increases with decreasing metallicity probably due to the
overlapping metallicity of stars from the halo, the thick-disk and the
thin-disk populations at this metallicity range.  Actually, the metallicity
gradient is more significant in the middle metallicity range of
$-1.2 < \feh < -0.4$ than the two outer ranges in this $\feh$ versus $|Z|$
diagram. 

In order to investigate the properties of the thick-disk population,
we try to exclude the contribution from the halo population
and plot in the lower panel of Figure~9 the $\feh$ versus $|Z|$ diagram for
stars with $|Z| < 5$ kpc and $\feh > -0.9$, which mainly originate from
the thick disk.  This figure shows that the trend in the $\feh$ versus $|Z|$
diagram is quite significant and the scatter around the gradient is greatly
reduced.  A linear regression to the data indicates a relation of $ \feh =
-0.292(\pm0.012) |Z| +0.116$ with a scatter of 0.037 dex.  Taking
into account the fact that a few points with  $\feh > -0.35$ do not
follow the above trend and may belong to the thin disk, they are excluded
from the fitting, and the result is $ \feh = -0.255(\pm0.014) |Z| +0.020$
with a scatter of 0.023 dex. The six points with  $\feh > -0.35$ follow
the trend of $\feh = -0.467(\pm0.083) |Z| +0.323 $ with a scatter of 0.031 dex
for $|Z| < 1.5$ kpc.

\begin{figure}[bt]
\includegraphics[scale=1.00]{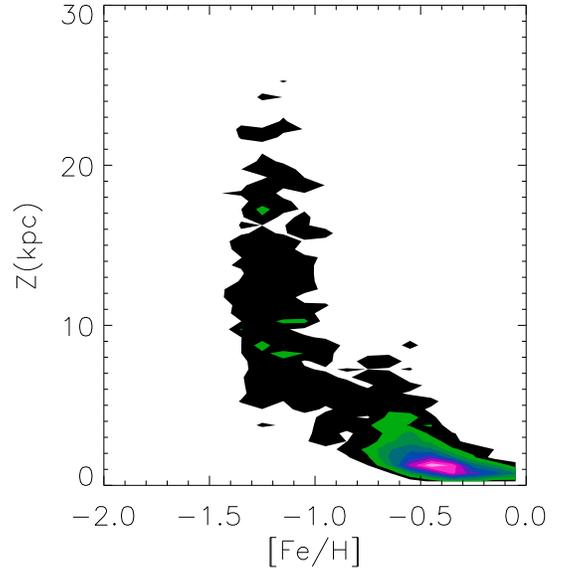}
\caption{Distribution of 5391 RHB stars in the $\feh$ vs. $|Z|$ diagram.
The star density decreases as the color varies from the pink, the blue, the green to
the black.}
\end{figure}

\begin{figure}[bt]
\includegraphics[scale=1.00]{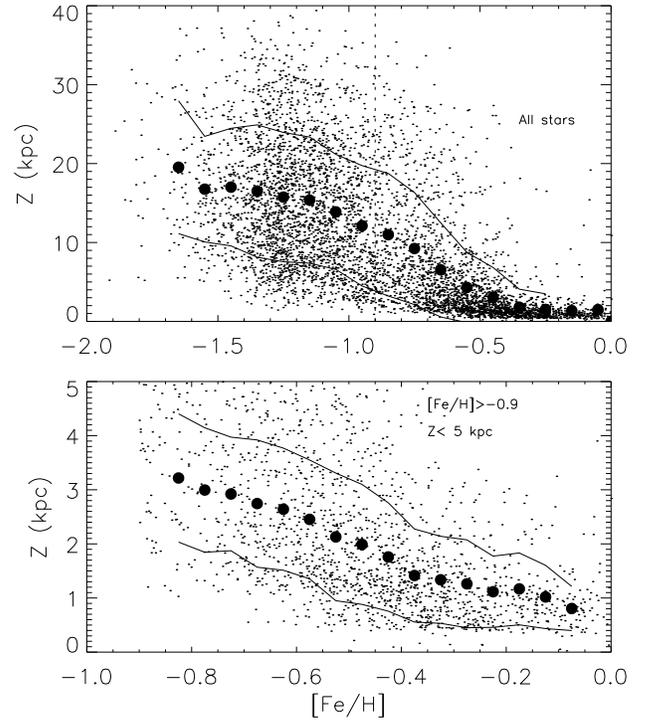}
\caption{$\feh$ vs. $|Z|$ diagram for the full sample (upper panel) and
for stars with $\feh > -0.9$ and $|Z| < 5$ kpc (lower panel). The mean $|Z|$
at a given $\feh$ and the scatter are shown by the large filled circles and
solid lines, respectively}
\end{figure}

\subsection{The $\Vrot$ versus $|Z|$ diagrams}

Figurere~10 shows the kinematical gradients, $\Vrot$ ($= V+220\,\kmprs$) versus
$|Z|$, and Figures~11 and 12 show the $U$ and $W$ velocities versus $|Z|$ for RHB
stars; for both figures,  $\feh < -0.9$ and $\feh > -0.9$ in the
upper and lower panels, respectively.  These figures show that ${\langle U \rangle}$
and ${\langle W \rangle}$ velocities generally do not vary much with $|Z|$, independent of
metallicity, while for $\feh < -0.9$ the ${\langle \Vrot \rangle}$ tends to decrease
with increasing $|Z|$ for the region $|Z|< 10 $ kpc and becomes flat for the region
$|Z| > 10$ kpc. The rotational-velocity gradients are approximately
$-30$ km s$^{-1}$ kpc$^{-1}$ for stars with $\feh <-0.9$ and $|Z|< 10 $ kpc, and
about $-19$ km s$^{-1}$ kpc$^{-1}$ for stars with $\feh >-0.9$ and $|Z|< 15 $ kpc.

Quantitatively, different authors give somewhat
different gradients for $|Z|< 5$ kpc.  For example, Carollo et al.\ (2010,
in their Figure~11) found a decreasing ${\langle V_{\phi} \rangle}$ with $|Z|$ for stars with
$-0.8 < \feh < -0.6$ and $|Z|< 4$ kpc with a rotational-lag gradient of
$-36$ km s$^{-1}$ kpc$^{-1}$.  Chiba \& Beers (2000) found a rotational-lag
gradient of $-30$ km s$^{-1}$ kpc$^{-1}$ based on a large sample of stars
($\sim 1200$) with $|Z| \la 2$ kpc.  Allende Prieto et al.\ (2006) adopted
SDSS photometric distances and radial velocities (without using proper motions),
and found a rotational-lag gradient of $-16$ km s$^{-1}$ kpc$^{-1}$ for stars
between $|Z| = 1$ and 3 kpc. In Majewski (1992), the value is $-21$ km s$^{-1}$ kpc$^{-1}$ 
based on deep proper-motion survey of the Galaxy out to $|Z| \sim 6$ kpc.
Our value of $-19$ km s$^{-1}$ kpc$^{-1}$ for stars with $\feh >-0.9$ and $|Z|< 15 $ kpc
is quite consistent with previous values within the error range.

It has been suggested that the
rotational lag and the velocity dispersions vary with distance from the
Galactic plane (Majewski 1994). However, our work shows that, for
$\feh < -0.9$, the ${\langle \Vrot \rangle}$ decreases with increasing $|Z|$ for the region
of $|Z|< 10 $ kpc, while for $|Z|> 10$ kpc, where the halo dominates, there is
no significant trend. In Bond et al.\ (2010) the halo population with
$\feh < -1.1$ does not show any trend in the ${\langle \Vrot \rangle}$ with $|Z|$ over the
region of $|Z|< 5 $ kpc, which does not agree with our result for $\feh < -0.9$
and $|Z|< 10 $ kpc.  This discrepancy may be easily understood if we assume that
stars with $\feh < -0.9$ in the present work consist of
two halo sub-populations.
Finally, the scatters around these trends in Figure.~10-12 are rather large,
and further studies with larger samples and high-quality data are needed.

\begin{figure}[bt]
\includegraphics[scale=1.00]{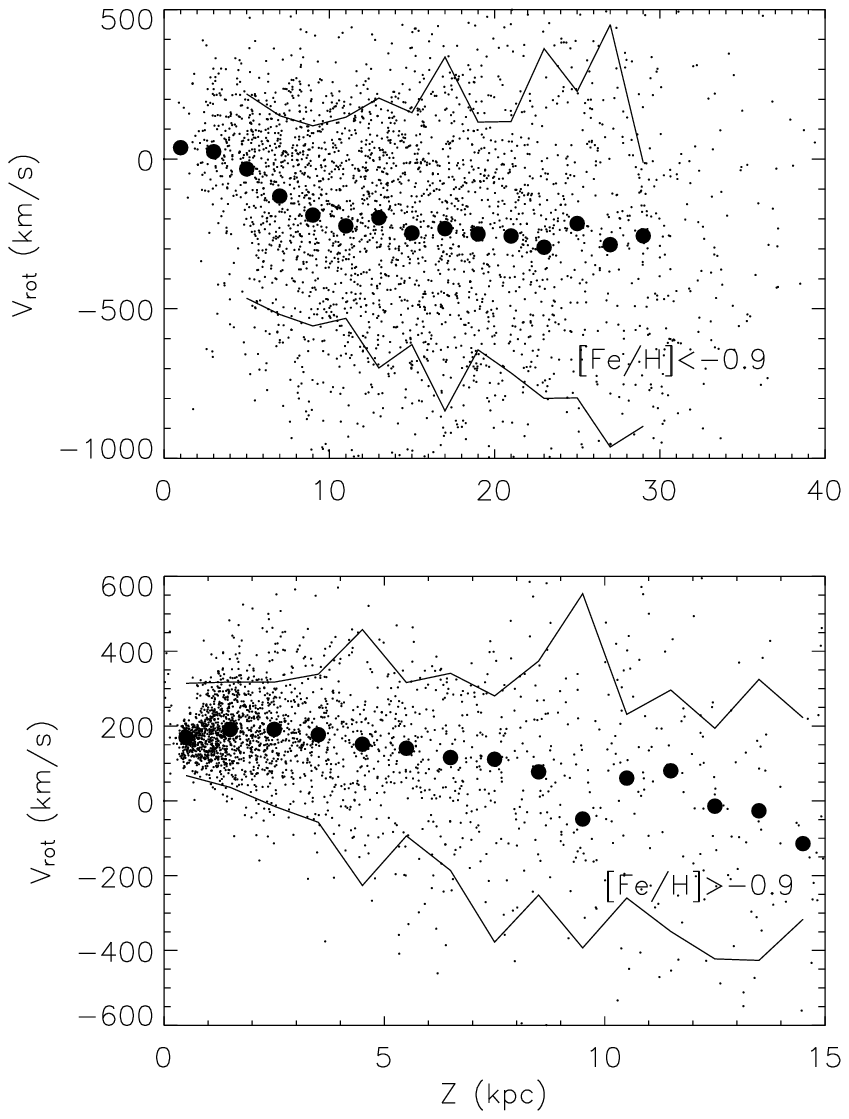}
\caption{$\Vrot$ versus $|Z|$ for stars with $\feh < -0.9$ and
$\feh > -0.9$ (upper and lower panels, respectively). }
\end{figure}

\begin{figure}[bt]
\includegraphics[scale=0.90]{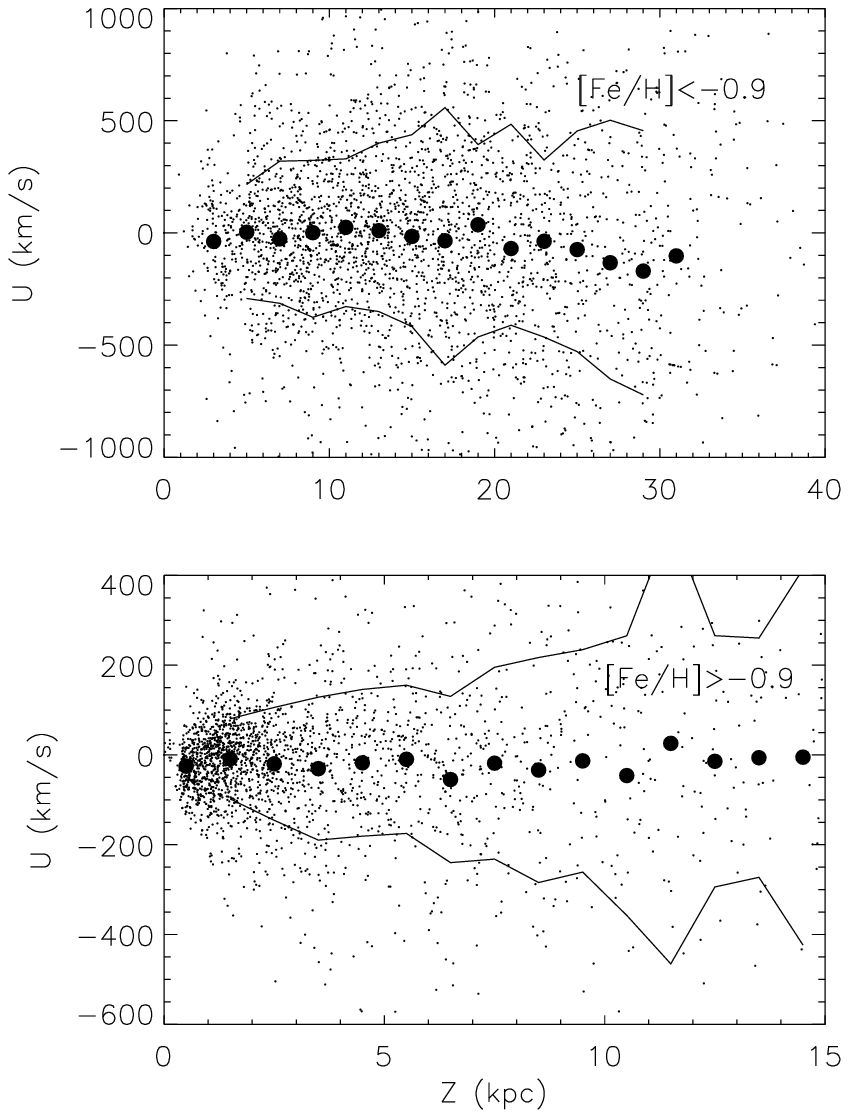}
\caption{$U$ velocities vs.
$|Z|$ for stars with $\feh < -0.9$ and $\feh > -0.9$ (upper and lower panels, respectively). }
\end{figure}

\begin{figure}[bt]
\includegraphics[scale=0.90]{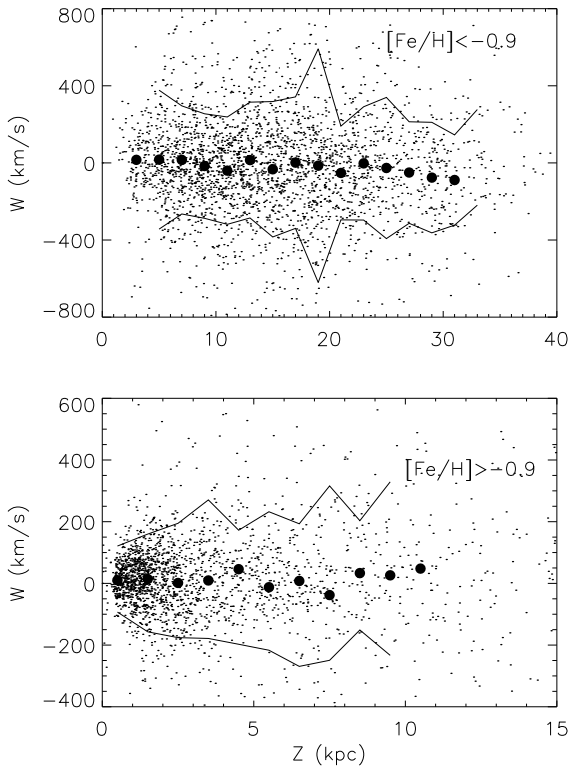}
\caption{$W$ velocities vs.
$|Z|$ for stars with $\feh < -0.9$ and $\feh > -0.9$ (upper and lower panels, respectively). }
\end{figure}

\subsection{The $\feh$ vs. $\Vrot$  diagram}

As described above, this diagram is a useful way to trace the structure
of the Galaxy when abundance and kinematical data are combined.  Figure~13
shows the histograms
of $\Vrot$ (upper panel) and the $\feh$ versus $\Vrot$ diagram (lower panel)
for RHB stars. The upper panel shows that
the metal-mild component peaks at $\Vrot \sim 170 \kmprs$
and the metal-poor component spans a wide range in $\Vrot$ without any sharp peak.
That is, the $\Vrot$ dispersion in the former is smaller than the latter component.
This is expected in the context of the Galactic stellar populations
with the halo population with $\feh < -1.0$ having a large velocity dispersion
as compared to the thick-disk population, with a metallicity peak
at $\feh \sim -0.6$, and the thin-disk population with a metallicity peak
at $\feh \sim -0.2$.
Actually, the distribution in $\Vrot$
for the metal-poor component is rather broad and 
in this work we separate the metal-poor component
into two groups with $\Vrot > 0 \kmprs$ and
$\Vrot < 0 \kmprs$ in the lower panel of Figure~13.
The metal-mild population with $\feh \sim -0.6$
has a peak at $\Vrot \sim 170 \kmprs$, which is quite similar to that derived from
the solar neighborhood by Soubiran et al.\ (2003), who derive a rotational lag
(Stromberg asymmetrical drift) of $V_{lag} \sim 51 \kmprs$ with respect to the
LSR, and is somewhat lower than the average rotational lag
of $V_{lag} \sim 80 \kmprs$ by Fuhrmann (1998) for a group of 16 thick-disk stars.
For the metal-poor population, we separate them into two sub-populations
with the division of stars with $\Vrot> 0 \kmprs$ being ``Halo I'' and
stars with $\Vrot< 0 \kmprs$ being ``Halo II''. 
In Figure~14, the $\feh$ versus $|Z|$ diagrams for the Halo I and II sub-populations
are shown in the upper and lower panels, respectively.  It shows that there is
a hint of metallicity gradient in the Halo I, as for the thick-disk
population, while for the Halo II population the metallicity gradient is much
weaker, if it exists at all.  This gives concordance to our separation of the
halo population into the Halo I and the Halo II components for stars with
$\feh < -0.9$ in Figure~13. 

Actually, the division of the halo at $\Vrot \sim 0 \kmprs$ is consistent
with the result of Carollo et al. (2007), who found an inner halo
with a slightly prograde rotation and a outer halo with a net retrograde.
In their Figure~5, the division between the inner and the outer
halo is quite clear in the $R - |Z|$ diagram. In view of this, it is interesting
to investigate the division for the halo population based on RHB stars. Figure~15 
shows the distribution of RHB stars for the Halo I and the Halo II 
sub-populations in the  $R - |Z|$ diagram. The main result from this figure
is that the Halo I stars clump within $R <10$ kpc and $|Z| < 10$ kpc,
while the Halo II stars show two clumps with one within 10 kpc
and the other grouping outside 10 kpc. In this sense, we favor the suggestion
that the transition between the inner halo and the outer halo is around
10 kpc. Note that the Figure~5 of Carollo et al. (2007) is divided by the
metallicity while our Figure~15 is separated by the rotation, and thus the
comparison between the two works is not direct. However, the structure
of the halo consisting the inner and the outer parts is the same.

\begin{figure}[bt]
\includegraphics[scale=0.90]{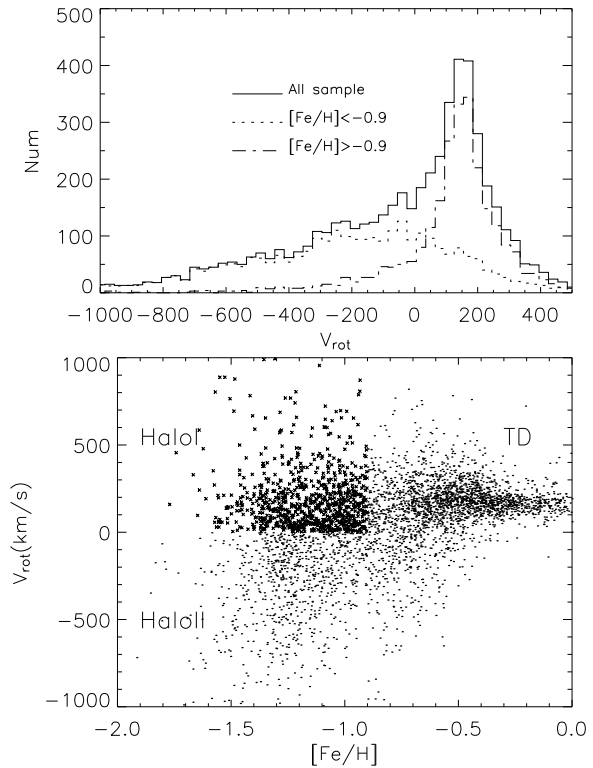}
\caption{Histograms of $\Vlsr$ (upper panel) and the $\feh$ versus $\Vlsr$
diagram (lower panel).  The thick-disk population consists of stars
with $\feh >-0.9$, while the halo population is divided into two sub-populations,
the Halo I (crosses) and the Halo II (dots).}
\end{figure}

\begin{figure}[bt]
\includegraphics[scale=1.00]{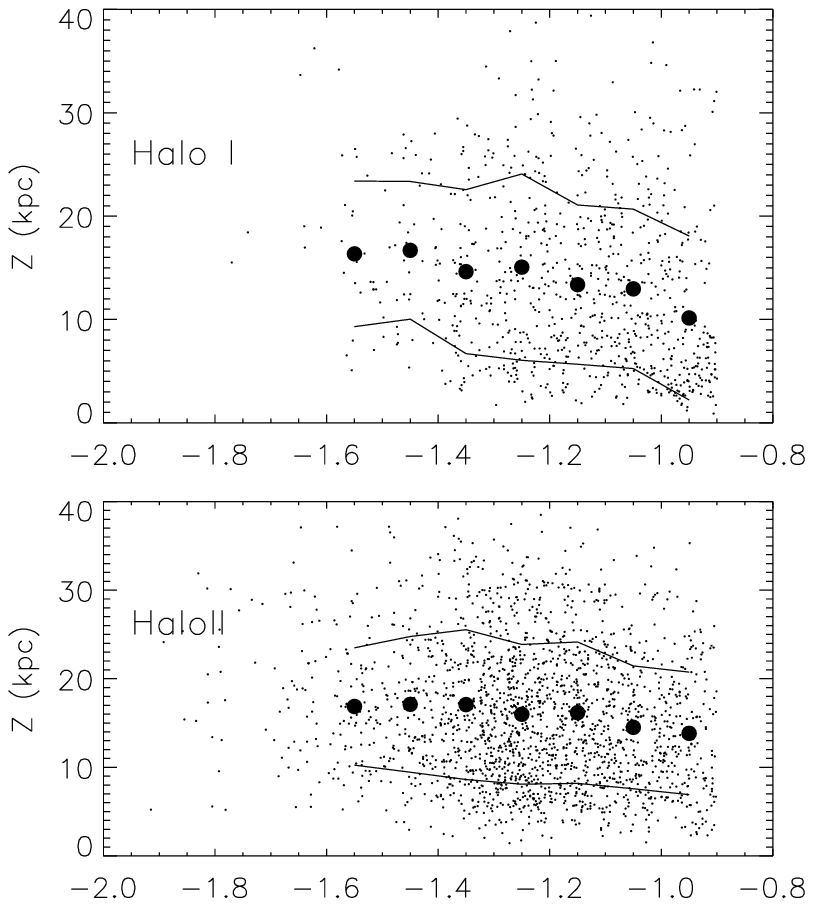}
\caption{$\feh$ vs. $|Z|$ for the Halo I (upper panel) and II (lower panel)
sub-populations.}
\end{figure}

\begin{figure}[bt]
\includegraphics[scale=1.00]{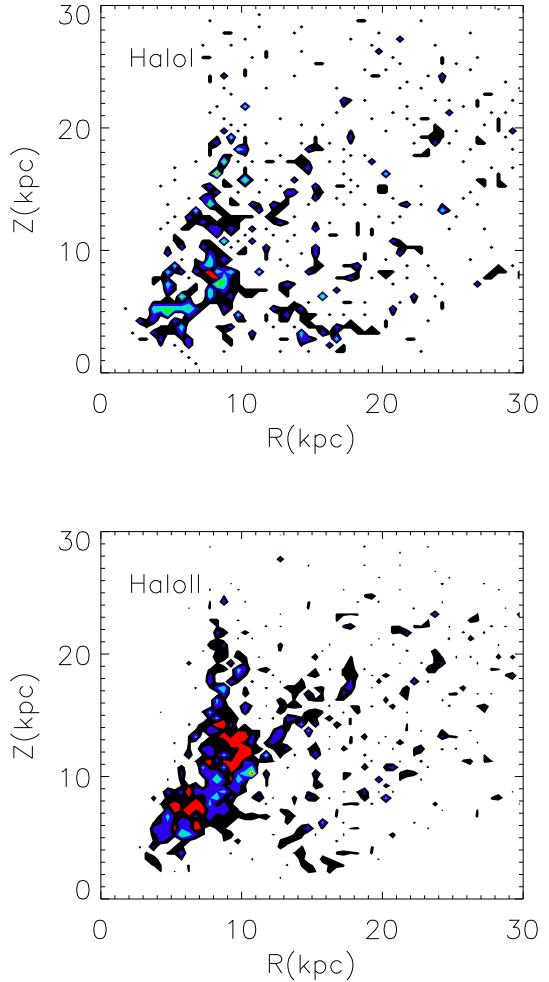}
\caption{$R$ vs. $|Z|$ diagrams for the Halo I (upper panel) and Halo II (lower panel)
sub-populations.}
\end{figure}

\subsection{Implications for the evolution of the Galaxy}

Since the first suggestion for the thick disk in our Galaxy was made
on the basis of star counts by Gilmore \& Reid (1983), a considerable
number of studies have helped to characterize the Galaxy's thick-disk
component, in an effort to deduce its origin and better understand its
nature.  Most works suggest that the thick disk consists of old
($\ga 10$ Gyr) stars of intermediate metallicities,
(-1.0 $\la$ [Fe/H] $\la$ -0.2) with a rotational lag of
$\approx 50$ km s$^{-1}$, and a scale height between 0.8 and 1.2 kpc.
Our results from RHB stars with $\feh < -0.9$ dex are generally
consistent with these numbers; this population peaks
at $\feh \sim -0.6\pm0.2$ dex,  $\Vrot \sim 170 \pm 20 \kmprs$,
and a vertical scale
height of $1.2\pm 0.3$ kpc.

However, the thick disk may have a metal-weak component (MWTD)
(e.g.,\ Beers et al.\ 2002).  Based on MS/TO stars from the SDSS survey, Ivezic et al.\
(2008) reveal a feature consisting of stars of $\feh \sim -1.0$ dex and
$|Z| \sim 2-3$ kpc, which they classified as the MWTD population.  Brown et
al.\ (2008) presented a complete spectroscopic survey of 2414 Two Micron All Sky Survey (2MASS)-selected
BHB candidates selected over 4300 deg$^2$ of the sky and
found that the BHB stars located at a distance from the Galactic plane $|Z|<4$
kpc trace what is clearly a metal-weak thick-disk population, with a mean
metallicity of [Fe/H] $=-1.7$ dex.  But the existence of MWTD among
BHB stars with $|Z| < 4$ kpc is 
not confirmed by their further work in the paper by Kinman et al.\ (2009).

In the present work, the MWTD has not been found as shown in Figure~14 (upper
panel), where there are only a few stars with $\feh < -0.9$ and $|Z|< 5$ kpc,
and very few with $|Z| < 2$ kpc. The Halo I sub-population with $\feh \sim -1.3$
and $\Vrot > 0 \kmprs$, generally has
$|Z| > 5$ kpc.  Since it is well accepted that the edge of the thick disk is at
about 5.5 kpc above the Galactic plane as suggested by Majewski (1994), and since
we have measured a scale height for the thick disk of $|Z|=1.2\pm 0.3$ kpc,
we suggest this sub-population, Halo I, really corresponds to the halo.

In connection with the halo population, our result confirms with the existence
of the inner and the outer halo sub-population in the $R$ - $|Z|$ diagram.
From the  $\feh$ - $|Z|$ diagram, it hints that
the inner halo shows a metallicity gradient while the outer
halo does not. From our data, we cannot detect any sign of stellar
streams reported before due to the selection effect of the spectroscopic
survey of the SDSS project and the star number in our sample is too
small for detecting any overdensity produced by stellar streams.

\section{Conclusions}
Based on photometric and spectral data, a group of RHB stars have
been selected from the SDSS survey, whose distances can be estimated
by using the calibrations from Chen et al.\ (2009).  Combining with
available proper motions from the USNO survey and radial velocities from the SDSS
survey, the space velocities (U,V,W) have been calculated for a sample
of 5391 RHB stars, and their metallicities and kinematical
gradients have been investigated in order to trace the evolution of the
Galaxy. The main results are as follows.  (1) There are two peaks in
the metallicity distributions of RHB stars with a division at
$\feh \sim -0.9$.  Stars with $\feh > -0.9$, centered at $\feh \sim -0.6$,
correspond to the thick-disk population, while stars with  $\feh < -0.9$,
peaking at $\feh \sim -1.3$, have generally halo kinematics in the Toomre
diagram. The metallicity gradient for thick disk stars with $-0.9 < \feh < -0.3$ is
significant with $\feh = -0.255|Z|+0.02$.  RHB stars from the thick disk
 with $|Z|<5$ kpc
have a peak metallicity of $\feh \sim -0.6$ dex, a peak
rotation velocity of $\Vrot \sim 170\,\kmprs$, and a
vertical scale height of $|Z| \sim 1.2$ kpc.  (2) The division between
the thick disk and the halo seems to be quite clear in
the $\feh$ versus $|Z|$ diagram as clearly shown in Figure 8, where the edge
of the thick disk could be as high as $|Z| \sim 8$ kpc at $\feh \sim
-0.6$ and reduces to $|Z| \sim 2$ kpc at $\feh \sim -1.5$. Meanwhile,
there is a detectable rotational gradient in the thick disk and
a nearly constant Galactic rotation in the halo in the vertical direction
to the Galactic plane.
(3) Two halo sub-populations
have been identified, the Halo I and the Halo II, for stars peaking at
$\feh \sim -1.3$,
based on the $\Vrot$ and the $\feh-|Z|$ diagram.  The Halo I mainly clumping at
$R < 10 $ kpc with a sign of metallicity  gradient, while
the Halo II clumping at two regions,  both $R < 10 $ kpc and $R > 10 $ kpc, shows a
negligible metallicity gradient.

The above results are somewhat limited by the presence of substantial
scatter around these relations, which probably derives mainly from the large
errors of the proper motions, and the limited sample of RHB stars with precise
and accurate proper motions.  Further study on a large sample of
RHB stars with accurate proper motions is desirable to clarify the two 
halo sub-populations. In particular, the distance and
proper-motion data for distant stars may be provided by the GAIA project.
Combining these with the upcoming LAMOST spectroscopic survey, we hope to
identify more RHB stars and to analyze the RHB sample for structure in the
metallicities, spatial coordinates, and Galactic velocities.

\acknowledgements
This work has been supported by the National Natural Science Foundation of China
under grants 10673015 and 10821061, the National Basic Research Program of
China (973 program) 2007CB815103/815403, the Academy program 2006AA01A120,
the Youth Foundation of National Astronomical Observatories of China, and the
CONACyT project CB-2005/49434 (Mexico).

   Funding for the Sloan Digital Sky Survey (SDSS) and SDSS-II
has been provided by the Alfred P. Sloan Foundation, the Participating Institutions,
the National Science Foundation, the US Department of Energy, the National Aeronautics
and Space Administration, the Japanese Monbukagakusho, the Max Planck Society, and the
Higher Education Funding Council for England. The SDSS Web
site is http://www.sdss.org. The SDSS is managed by the Astrophysical Research
Consortium (ARC) for the Participating Institutions. The Participating
Institutions are the American Museum
of Natural History, the Astrophysical Institute Potsdam, the
University of Basel, the University of Cambridge, Case Western
Reserve University, the University of Chicago, Drexel University,
Fermilab, the Institute for Advanced Study, the Japan Participation Group,
The Johns Hopkins University, the Joint Institute for
Nuclear Astrophysics, the Kavli Institute for Particle Astrophysics
and Cosmology, the Korean Scientist Group, the Chinese Academy of Sciences (LAMOST),
Los Alamos National Laboratory,
the Max Planck Institute for Astronomy (MPIA), the Max Planck
Institute for Astrophysics (MPA), New Mexico State University,
Ohio State University, the University of Pittsburgh, the University
of Portsmouth, Princeton University, the United States Naval Observatory,
and the University of Washington.

\end{document}